\documentclass[sigconf]{acmart}
\usepackage{amsthm}
\newtheoremstyle{user}
  {2pt}   
  {2pt}   
  {\normalfont}
  {0pt}
  {\bfseries}
  {}
  {0pt}
  {\thmnumber{(#2) }}
\theoremstyle{user}
\newtheorem{prop}{}

\AtBeginDocument{%
  }

\setcopyright{acmlicensed}
\copyrightyear{2026}
\acmYear{2026}
\acmDOI{XXXXXXX.XXXXXXX}
\acmConference[ACM CAIS 2026]{ACM Conference on AI and Agentic Systems}{May 26--29, 2026}{San Jose, CA}
\acmISBN{978-1-4503-XXXX-X/2026/06}

\begin{document}

\title[Querying Everything Everywhere All at Once]{Querying Everything Everywhere All at Once:\\Supervaluationism for the Agentic Lakehouse}

\author{Jacopo Tagliabue}
\email{jacopo.tagliabue@bauplanlabs.com}
\affiliation{%
  \institution{Bauplan Labs}
  \city{New York}
  \country{USA}
}

\renewcommand{\shortauthors}{Tagliabue et al.}

\begin{abstract}
    Agentic analytics is turning the lakehouse into a multi-version system: swarms of (human or AI) producers materialize competing pipelines in data branches, while (human or AI) consumers need answers without knowing the underlying data life-cycle. We demonstrate a new system that answers questions \emph{across} branches rather than \emph{at} a single snapshot. Our prototype focuses on a novel query path that evaluates queries under supervaluationary semantics. In the absence of comparable multi-branch querying capabilities in mainstream OLAP systems, we open source the demo code as a concrete baseline for the OLAP community.
    \end{abstract}


\begin{teaserfigure}
  \includegraphics[width=\textwidth]{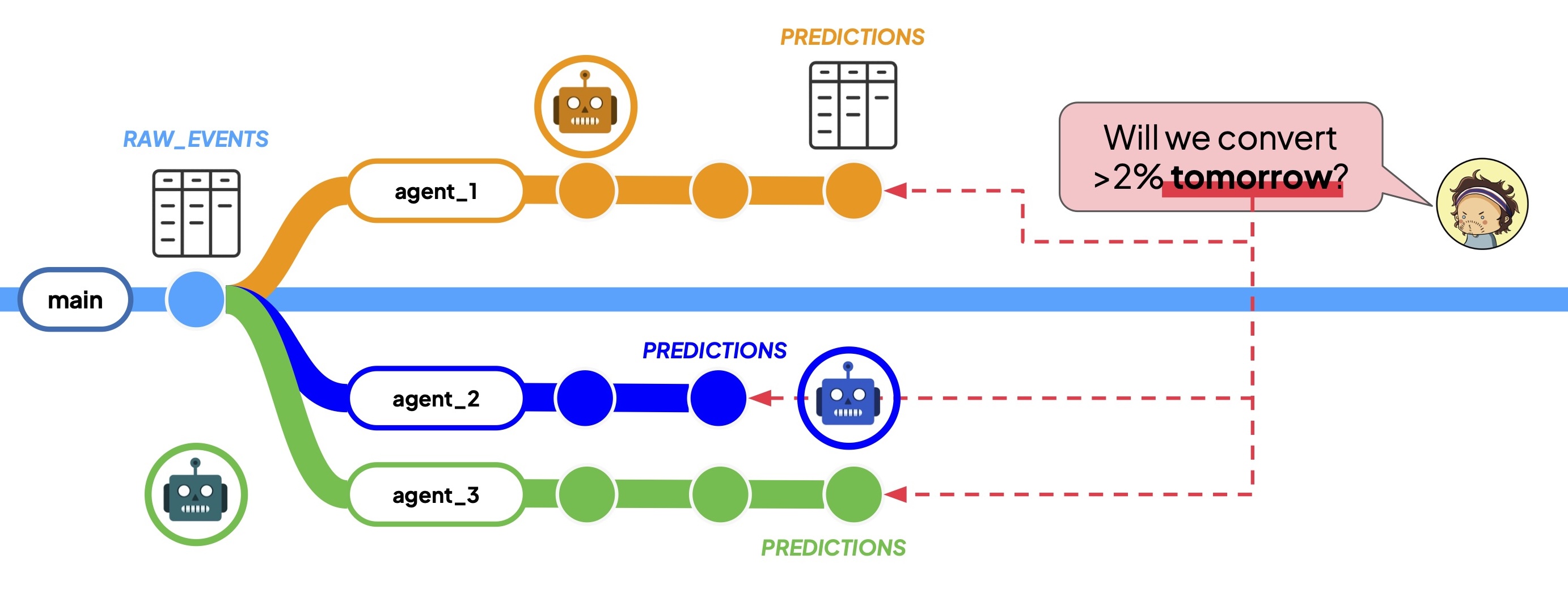}
  \caption{\textit{ACME Inc.} demo scenario: three agentic pipelines run concurrently on lakehouse branches, producing divergent versions of analytical tables. When a user asks a question, can we answer it without assuming a canonical data representation?}
  \label{fig:teaser}
\end{teaserfigure}


\maketitle

\section{Introduction}

\begin{quotation}
``Falsehood is never in words; it is in things.'' \\
Italo Calvino, \textit{Invisible Cities}
\end{quotation}

In the year of the (over)lord(s)~\cite{liu2025supportingaioverlordsredesigning} \textit{one}, \textit{ACME Inc.} is an online shop using agents for reporting and forecasting; e.g., a data engineer kicks off a purchase prediction job, resulting in three agents swarming as in Fig.~\ref{fig:teaser}. What if a user asks about tomorrow's sales, blissfully unaware of the data life-cycle? All versions of the forecast agree that \textit{some} sales will happen: the conversion rate is \textit{definitely} \textgreater{}0. But in some versions \textgreater{}2\% of shoppers end up buying, in some less.

Data analyses produced independently by multiple agents often disagree on some facts while sharing common ground. For example, two agents computing monthly revenue may disagree on the exact figure, yet both agree that revenue grew quarter-over-quarter. We argue that a non-classical logic, \textit{supervaluationism}, applies naturally to this setting: rather than waiting for, or forcing, a single canonical version, we reason \textit{across} the versions. The key insight is that even when there is no lakehouse-wide agreed-upon version of data assets, we can still draw reliable conclusions in many cases \cite{Varzi1997-VARIWC}. 

In particular, we demonstrate an AI system on top of a cloud lakehouse \cite{Zaharia2021LakehouseAN} that answers user questions across data versions, instead of the per-version querying of standard OLAP engines. Time travel answers ``query at snapshot $t$''; in contrast, we answer ``query across all  branches''. We summarize our contributions as follows:

\begin{enumerate}
    \item we ground our use case in running an agentic lakehouse at scale, where every week hundreds of thousands of \textit{data branches} \cite{tagliabue2025safeuntrustedproofcarryingai} are created in production;
    \item we provide definitions for the relevant semantics both intuitively and formally, and showcase the difference between an \textit{ad hoc} and \textit{native} implementation;
    \item we release the entire demo\footnote{\textbf{Video}: \url{https://youtu.be/REPLACE_WITH_PRIVATE_LINK}.} as open source code under a permissive license\footnote{Available at: \url{https://github.com/BauplanLabs/querying-everything-everywhere-all-at-once}. \textbf{Note for reviewers}: the demo is fully reproducible, but requires access to cloud services such as \textit{OpenAI} and \textit{bauplan} -- feel free to reach out for access. A lightweight \textit{replay mode} over exported Parquet files is also available to test the engines without any cloud dependencies.}.
\end{enumerate}

This demonstration showcases \textit{technical} innovation motivated by \textit{interactive} use. As such, we believe it to be interesting to a wide set of practitioners, as it addresses a \textit{full vertical slice} for a high-value agentic scenario: from business user needs and organizational constraints, down to system-level optimizations inside a query engine. Furthermore, at the moment of writing \textit{this} paper, no major \textit{OLAP} platform\footnote{As for example the top three lakehouses, \textit{Databricks}, \textit{Snowflake}, and \textit{Fabric}. A few OLTP solutions exist, but obviously they cover different use cases than data pipelines and data science. } has branching APIs nor even support for arbitrary transactions, making direct comparisons impossible: by releasing code grounded in industry open standards (e.g. \textit{D3.js}, \textit{Iceberg} \cite{iceberg}, \textit{DataFusion}\cite{10.1145/3626246.3653368}), we hope to foster the community interest in agent-first architectures. Finally, mostly as a nod to our younger selves, we are delighted to observe that conflicting data sources were the motivating examples for Belnap's seminal work \cite{Belnap1977} in multi-valued logic.\footnote{``The computer is to be envisioned as obtaining the data (...) from a variety of sources, all of which indeed may be supposed to be on the whole trustworthy, but none of which can be assumed to be that paragon of paragons, a universal truth-teller''.}

\section{Motivation}

The rise of AI agents and patterns such as \textit{Ralph} \cite{ralph} and \textit{Self-Driving Codebases} \cite{cursor} highlight a novel challenge for OLAP systems: instead of a slow-moving code base managed by a few, trusted data engineers and scientists working in sequence, we now have massively parallel experimentation done by a swarm of untrusted agents \cite{liu2025supportingaioverlordsredesigning}. As experimental code lives in code branches, \textbf{data producers} should materialize experimental data assets in \textit{data branches}, i.e., copy-on-write versions of the lakehouse with strong isolation guarantees \cite{sheng2026buildingcorrectbydesignlakehousedata}. Through the privileged lens provided by \textit{bauplan} \cite{10.1145/3702634.3702955} traces across its deployments -- with over a million branches created in the last quarter -- we observe that at any given point in time, it is now \textit{likely} that multiple versions of the same table coexist (Fig.~\ref{fig:teaser}): data pipelines may be long-running processes and not yet completed; or, pipelines may have been completed, but no PR has happened yet; or, alternative answers should coexist, as the original intent was genuinely polysemic.

This is not the only acceleration happening. On the \textbf{data consumer} side, the slow ``business to analyst to engineering'' handover is being quickly dismantled by text-to-SQL and data agents \cite{hong2025nextgenerationdatabaseinterfacessurvey}. The user experience quickly degrades, however, as questions are asked on tables existing in multiple active branches: as the business user has neither the knowledge nore the access to move forward, she is now once again slowed down by processes and infrastructure she doesn't understand.

\textit{This} system is a demonstration that another way is possible: (human or agentic) data producers and (human or agentic) data consumers can be less tightly coupled if only we could reason consistently across multiple versions of the same tables. In other words, if only we could query \textit{everything everywhere all at once}.

\section{System Overview}

We start from a cloud lakehouse with Git-like primitives over Iceberg tables: \textit{commits}, \textit{branches}, and \textit{merges} (see \cite{10.1145/3650203.3663335,Tagliabue2023BuildingAS,tagliabue2025trustworthyaiagenticlakehouse}). The core of our demonstration is the \textit{query flow}, i.e. going from English questions to answers while reasoning across branches. 

\subsection{Architecture}
\begin{figure}
    \centering
    \includegraphics[width=\columnwidth]{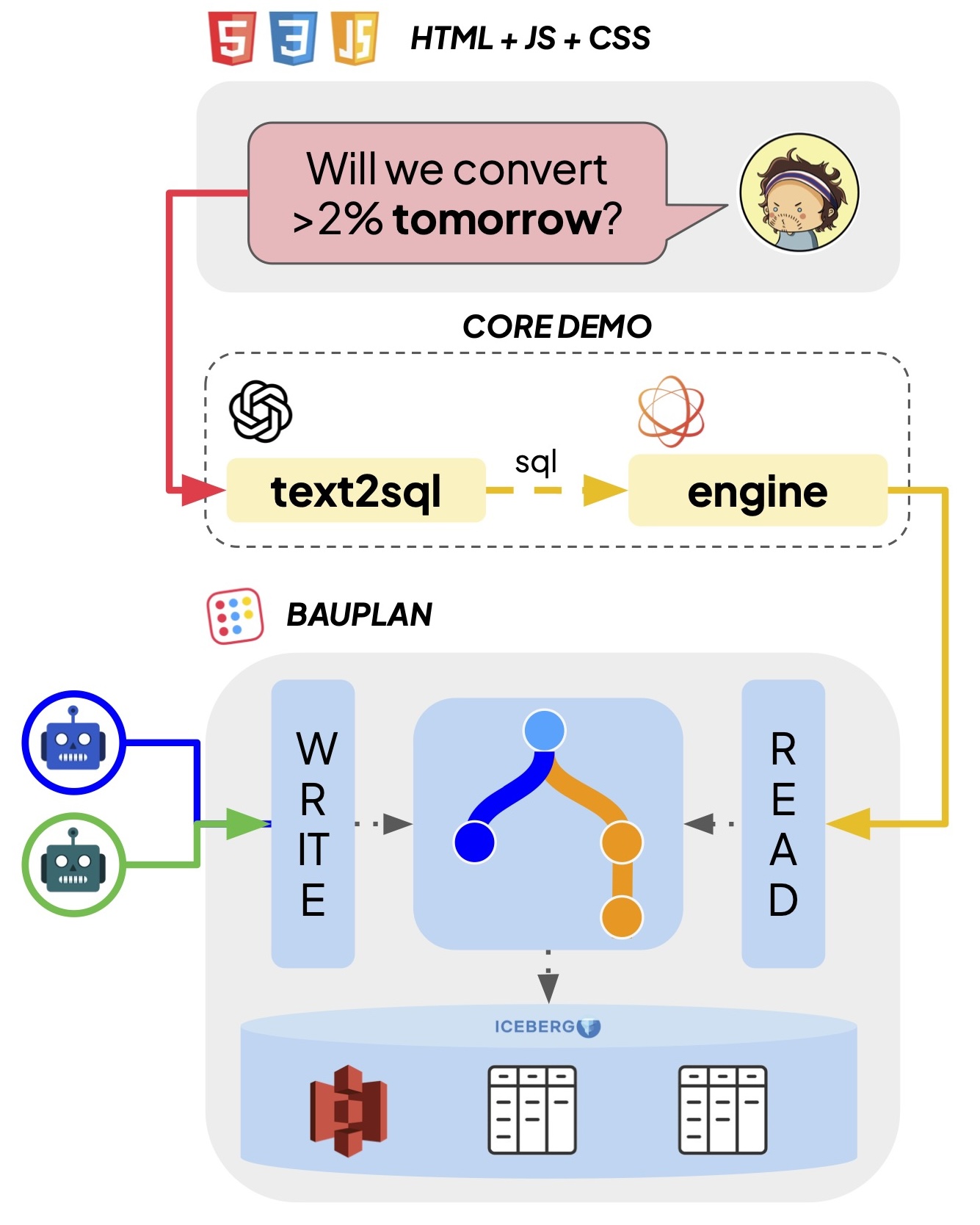}
    \caption{Architecture of the demo system. Agentic pipelines write predictions into isolated \textit{data branches} of the lakehouse. The UI translates English questions into SQL, then executes them using a modified OLAP engine that evaluates answers across branches under supervaluationary (glut-tolerant) semantics.}
    \label{fig:arch}
\end{figure}

The architecture is fairly straightforward (Fig.~\ref{fig:arch}). At the bottom, a branching lakehouse allows data agents to write data pipelines concurrently (Fig.~\ref{fig:teaser}); at the top, a \textit{D3-based app} provides a GUI, backed by a \textit{FastAPI} server. The core contribution of this demo \textit{is the translation module in the middle}: an LLM-powered text-to-SQL translation layer and a modified DataFusion OLAP engine, implementing the supervaluationist semantics (Section~\ref{sec:superval}) when querying data branches in the lakehouse.

Our goal for the text-to-SQL sub-module is to support the demo experience: based on our testing, the narrow domain makes the task trivial for off-the-shelf models, so we claim no contribution here on top of implementing best practices from the recent literature \cite{hong2025nextgenerationdatabaseinterfacessurvey}, such as schema augmentation and feedback-driven regeneration, i.e. trying to produce query plans and fail early instead of running a query directly from an LLM completion.

\subsection{Supervaluationism}
\label{sec:superval}

We first provide an explanation of non-classical semantics applied to reasoning across branches. We then describe the implementation inside a state-of-the-art engine and state known limitations and directions for future work.

\subsubsection{Logic} In our scenario (Fig.~\ref{fig:teaser}), \textit{ACME Inc.} is an e-commerce company where data agents are tasked to compute (possibly competing) predictions on future shoppers' behavior. Consider the following statements:

\begin{prop}\label{puno}
Conversion is \textgreater{}2\%.
\end{prop}
\begin{prop}\label{pdue}
It is not the case that conversion is \textgreater{}2\%.
\end{prop}
\begin{prop}\label{ptre}
Conversion is \textgreater{}2\% and it is not the case that conversion is \textgreater{}2\%.
\end{prop}
\begin{prop}\label{pfour}
Conversion is \textgreater{}2\% OR revenues are non-negative.
\end{prop}

\ref{puno} is true \textit{in some branch}, \ref{pdue} \textit{in some (other) branch}, but \ref{ptre} is never true: \textit{in some branch} does not pass through the (truth-functional) conjunction \cite{05b0adc3-0618-37f6-9323-153e114e4efa}. Notably, not anything goes, as we still have enough logical structure to be able to assert \ref{pfour}, which is true even if half of it is only true \textit{in some branch}. In a \textit{supervaluationary}\footnote{Since we are dealing with truth gluts and not gaps, ``subvaluationary'' is technically more appropriate: with a slight abuse of notation, we choose to use the more familiar term throughout the paper.} approach, \ref{puno} is a truth \textit{glut} \cite{Varzi1997-VARIWC}, a statement to which the model\footnote{``Model'' in the sense of model theory, not AI: we rely on the context to disambiguate.} assigns \textit{multiple} truth values (which is classically impossible). At evaluation time, if we get the same outcome no matter how we ``clear the gluts'', then the gluts don’t matter and sentences may be true (false) \textit{simpliciter}. For example, assuming revenues are non-negative, \ref{pfour} is true in any ``contraction'' of the model, as it doesn't matter how a particular branch clears the glut of the first disjunct. Armed with this core intuition, we now tackle the engineering challenges: we refer readers to Appendix~\ref{sec:formal} for a formal sketch.

\subsubsection{Engineering}
As supervaluationary semantic assignments are parasitic over per-branch assignments, a first implementation strategy --- which we call the \textit{ad hoc} engine, as it is bolted on top of an existing engine --- suggests itself: evaluate the query $Q$ independently on each branch and concatenate the per-branch outputs, tagging each row with a branch identifier column. For instance, ``How many buyers do we expect tomorrow?'' becomes a \texttt{SELECT COUNT(*)} executed once per branch, producing one scalar result per branch.

Table~\ref{tab:multiverse-questions} contains common business questions for our \textit{ACME Inc.} demo scenario \cite{roychowdhury2020categorizingonlineshoppingbehavior}. We divide them by the result \textbf{type} --- number, boolean, or list --- which determines the outer semantic logic: numbers are shown per branch with a compact summary; booleans yield a supervaluationary verdict (true, false, or mixed); lists are compared via Git-style \textit{diffs}.

\smallskip
\noindent\textbf{Native multi-branch planning.}
There are reasons to be dissatisfied with this ad hoc approach. First, answering questions such as $Q2$ in the demo involves a \texttt{JOIN}, and the ad hoc strategy repeats the same work $B$ times even when some inputs are identical across branches. More subtly, the supervaluationary approach uncovers optimizations that depend on the result type. We therefore implement a \textit{native} multi-branch engine that dispatches to specialized execution paths.

For aggregate and join queries (e.g. $Q2$-$Q3$), the native engine uses a custom DataFusion \texttt{TableProvider} that exposes all branch variants of a table as a single virtual relation with an injected \texttt{\_\_branch\_id} column. The entire query plan runs in a single Rust process, giving the optimizer full visibility for predicate pushdown, projection pruning, and shared-subplan reuse.

For boolean queries such as $Q4$, the native engine also supports a short-circuit evaluator: it runs per-branch evaluations in parallel and terminates as soon as two branches disagree, since the verdict (mixed) is then determined without scanning the remaining branches. Such early termination is difficult to express as static SQL rewriting, but natural to implement in the multi-branch coordinator. We measure the performance gap between the ad hoc and native strategies in Section~\ref{sec:experiments}.

\subsubsection{Limitations and future work}
The system defines supervaluationary semantics over three target question types, and provides a small set of initial benchmarks (Section~\ref{sec:experiments}) to start mapping out the design space. The approach can be generalized in several ways. \textbf{At the engine level}, we could push evaluation for more complex, non-scalar queries. \textbf{At the user level}, a similar approach could work for on-the-fly analytical jobs in response to polysemic requests; e.g., when reporting on ``customer lifetime value'', we could reason over multiple precisifications \cite{Pinkal1995} of that concept. \textbf{At the formal level}, a complete treatment would model inter-branch relationship: e.g., commit histories could form a modal structure with truth evaluated locally at a world \cite{Priest_2008}. Broadly speaking, our north star (and perhaps the most valuable insights of this work) is to be able to treat ``quarantined  inconsistencies'' \cite{05b0adc3-0618-37f6-9323-153e114e4efa} and ``conceptual indeterminacy'' \cite{lvarez2018DealingWC} as first-class citizens of our interactions with data systems, instead of something we need to ``fix'' \textit{before} analysis can begin. 

Aside from technical contributions, the system effectively defines a novel API for agent-first OLAP systems. Instead of querying tables \texttt{AT A SNAPSHOT}, we are moving up in abstraction by querying \textit{the entire lakehouse all at once}. As the deployments of swarms of data agents rise \cite{liu2025supportingaioverlordsredesigning,shapira2026agentschaos} and English becomes the primary way of interacting with our data, we may benefit from systems that can reason in a principled way over truth gaps and truth gluts alike.

\section{Demonstration scenarios}

We structure the demonstration as a progressive reveal of the new query abstraction: in the \textit{ACME Inc.} scenario, a user asks an English question, the system compiles it to SQL, executes it across data branches, and returns an answer whose format depends on the type.

\smallskip
\noindent\textbf{Phase 1 (setup and branching).}
We run $k$ pipelines (pre-generated with ChatGPT 5.2) in parallel over \textit{ACME Inc.} data \cite{kaggle}, materializing their outputs into separate lakehouse branches. The UI visualizes the evolving branch graph, making multiple data versions as concrete as Git.

\smallskip
\noindent\textbf{Phase 2: ``How many customers will buy tomorrow?''}
We start with a scalar aggregation ($Q1$), a count over the \texttt{predictions} table. The UI shows one number per branch plus a compact summary, establishing that pipelines can disagree on the same business question.

\smallskip
\noindent\textbf{Phase 3: ``Is tomorrow's conversion rate above 2\%?''} We move to a boolean threshold query ($Q4$). The UI reports per-branch true/false outcomes and a final verdict.

\smallskip
\noindent\textbf{Phase 4: ``How many smartphone shoppers will convert tomorrow?''} We introduce a cross-table join ($Q2$): the query filters a shared 3M-row user-dimension table by interest (i.e., smartphones) and joins it with the per-branch predictions table. User data are stable across branches while predictions diverge, highlighting queries where some inputs are shared and others are branch-specific.

\smallskip
\noindent\textbf{Phase 5: ``Which customers should we target tomorrow?''} Finally, we ask a set-valued question ($Q6$). Each branch returns its own top-$k$ user IDs; the UI displays a git-diff-style view of the consensus.

\smallskip
\noindent\textbf{Variations.}
While we provide a scripted progression of questions, the UI is not hard-coded to those and can potentially answer arbitrary questions from the people in attendance (subject to the schema and supported types).

\begin{table}
  \caption{Sample questions by result type}
  \label{tab:multiverse-questions}
  \begin{tabular}{cp{0.60\linewidth}c}
    \toprule
    ID & User-facing question & Result type \\
    \midrule
    Q1  & How many customers are expected to buy tomorrow? & \textbf{Number} \\
    Q2  & How many smartphone shoppers will convert tomorrow? &  \\
    Q3  & How many customers in the top segments are expected to buy tomorrow? &  \\
    \midrule
    Q4  & Is tomorrow's conversion rate above $\tau$? & \textbf{Boolean} \\
    Q5  & Is customer $u$ expected to buy tomorrow? &  \\
    \midrule
    Q6 & Which customers are expected to buy tomorrow? & \textbf{List} \\
    Q7 & Which customers are undecided and might be influenced by messaging? &  \\
    \bottomrule
  \end{tabular}
\end{table}

\section{Experiments}
\label{sec:experiments}

To support the system design, we report a small set of targeted experiments on representative query types. Our goal is not exhaustive benchmarking, but to show that native multi-branch execution already yields measurable benefits for typical agentic access patterns  \cite{liu2025supportingaioverlordsredesigning}. We pick representative queries, and run them over test tables varying the number of branches $B \in \{1,\ldots,50\}$ (Appendix~\ref{sec:further-experiments}).

\begin{figure}
    \centering
    \includegraphics[width=\columnwidth]{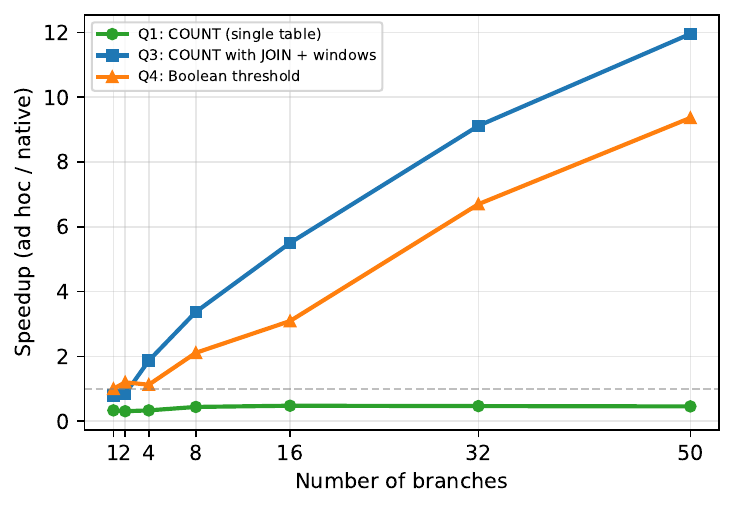}
    \caption{Speedup of the native engine over the ad hoc strategy as a function of branch count. The JOIN query with window functions (Q3) benefits from shared-table reuse; the boolean query (Q4) achieves near-constant time via short-circuit evaluation.}
    \label{fig:bench-speedup}
\end{figure}

Fig.~\ref{fig:bench-speedup} shows the speedup. For simple single-table counts (Q1), the unified plan's coordination overhead outweighs the benefit of a single plan, resulting in a $0.5{\times}$ ratio at $B{=}50$. For Q3 --- a JOIN with window functions over the shared dimension table --- the picture reverses: the ad hoc engine recomputes the expensive CTE per branch, while the native engine computes it once and reuses the hash-join, yielding $12.0{\times}$ at $B{=}50$. For boolean queries, the supervaluationary semantics enables a short-circuit optimization: once two branches disagree the verdict is determined, so Q4 stays flat at ${\sim}11$\,ms regardless of $B$.

\section{Related work}
\smallskip
\noindent\textbf{Database branching.}
Git-like semantics for data have been explored before (for humans): Dolt \cite{dolt} supports SQL only (OLTP, not OLAP), while Nessie \cite{nessie} is only a catalog, without pipeline APIs. Industry lakehouses built on Iceberg per-table evolution support \texttt{QUERY AT t}, but lack the API expressivity to formulate our question. Building on a system supporting transactions across tables, languages, operations \cite{sheng2026buildingcorrectbydesignlakehousedata} unlocks a richer UX, backed by a genuinely novel query semantics.

\smallskip
\noindent\textbf{Non-classical logic.}
The basic insights for reasoning over ``precisifications'' come from supervaluationism \cite{Pinkal1995}, which is easily adapted to gluts in the case of paraconsistency \cite{Varzi1997-VARIWC}. Our contribution is in mapping the theoretical semantics onto an agentic lakehouse at industry scale, yielding a new end-to-end experience for agentic analytics.

\bibliographystyle{ACM-Reference-Format}
\bibliography{sample-base}

\appendix

\section{AI Disclaimer}
The companion repository was built with a combination of AI coding assistants: ChatGPT 5.2 for planning, Claude Code with Opus 4.6 for implementation, and roborev \cite{roborev} for adversarial code reviews (through Codex). This paper was, however, entirely written by humans (aside from typos, proofreading, and similar minor edits).

\section{Supervaluationism: formal remarks}
\label{sec:formal}

We give a model-theoretic sketch tailored to the concrete KPI-style sentences used throughout the demo. Recall that the motivating intuition is that a sentence such as $\mathsf{Revenue} > X$ may be supported by one branch and refuted by another. Following Varzi's, a sentence is true (false) in an inconsistent model iff it is true (false) in some consistent contraction of that model \cite{Varzi1997-VARIWC}. In our setting, ``contractions'' are the per-branch classical evaluations, and we will sketch how to model that KPI sentence in a small FOL fragment.

\paragraph{Language.}
Fix a quantifier-free KPI language $\mathcal{L}_{\mathrm{kpi}}$ consisting of:
(i) constant symbols such as $\mathsf{Revenue}$,
(ii) a constant symbol $\bar{X}$ for each numeric threshold $X$ of interest, and (iii) binary predicate symbols $>$ and $=$. We are interested in sentences of the form
\[
\varphi_X \;:=\; (\mathsf{Revenue} > \bar{X}).
\]
\paragraph{Branch models.}
Each branch $b$ induces a classical $\mathcal{L}_{\mathrm{kpi}}$-structure $\mathcal{M}_b$ with domain $|\mathcal{M}_b|=\mathbb{R}$ and standard interpretations of comparison predicates:
\[
>^{\mathcal{M}_b}=\{(a,c)\in\mathbb{R}^2 \mid a>c\},
\qquad
=^{\mathcal{M}_b}=\{(a,c)\in\mathbb{R}^2 \mid a=c\}.
\]
Each KPI is interpreted by evaluating the corresponding query on branch $b$ (e.g. $\mathsf{Revenue}$ evaluates to a real number). Satisfaction is then classical:
\[
\mathcal{M}_b \models (\mathsf{Revenue} > \bar{X})
\;\;\Longleftrightarrow\;\;
\mathsf{Revenue}^{\mathcal{M}_b} > X.
\]
\paragraph{An inconsistent model via contractions.}
Rather than defining a single non-classical numeric structure directly, we model the lakehouse state by the set of its admissible \emph{consistent contractions}. Formally, let $\mathcal{M}$ be an ``inconsistent'' lakehouse whose contractions are the per-branch classical models:
\[
\mathrm{Con}(\mathcal{M}) \;:=\; \{\mathcal{M}_b \mid b\in B\}.
\]
Intuitively, each $\mathcal{M}_b$ corresponds to one way of ``clearing'' the disagreement by committing to the values observed on a particular branch.

\paragraph{Subvaluational truth conditions (gluts).}
For any sentence $\psi$ of $\mathcal{L}_{\mathrm{kpi}}$, define truth (and, similarly, falsity) in the lakehouse $\mathcal{M}$ by quantification over contractions:
\[
\mathcal{M} \vDash^+ \psi
\;\;\stackrel{\mathrm{def}}{\Longleftrightarrow}\;\;
\exists \mathcal{N}\in\mathrm{Con}(\mathcal{M})\;(\mathcal{N}\models \psi)
\]
Thus a sentence can be both true and false in $\mathcal{M}$ whenever different branches disagree, but that doesn't mean that ``anything goes'', as we demonstrate below.

\paragraph{Worked examples.}
Fix $X=100$ and consider the KPI sentence $\varphi := (\mathsf{Revenue}>\bar{100})$.

\smallskip
\noindent\textbf{(Glut.)}
Suppose there are two branches $a,b\in B$ such that
\[
\mathsf{Revenue}^{\mathcal{M}_a}=120,
\qquad
\mathsf{Revenue}^{\mathcal{M}_b}=80.
\]
Then $\mathcal{M}_a\models\varphi$ while  $\mathcal{M}_b\not\models\varphi$.

By the subvaluational truth conditions,
\[
\mathcal{M}\vDash^+ \varphi \quad\text{and}\quad \mathcal{M}\vDash^- \varphi,
\]
so $\varphi$ is both true and false in $\mathcal{M}$: a truth glut driven by branch disagreement.

\smallskip
\noindent\textbf{(Definitely true.)}
Reuse the same two branches $a,b\in B$, but let
\[
\varphi' \;:=\; (\mathsf{Revenue}>\bar{50}).
\]
Then $\mathcal{M}_a \models \varphi'$ and $\mathcal{M}_b \models \varphi'$ since $120>50$ and $80>50$. Consequently,
\[
\mathcal{M}\vDash^+ \varphi'.
\]

Thus $\varphi'$ is \emph{definitely true} across the lakehouse: although branches may disagree about other thresholds, for $X'=50$ the statement $\mathsf{Revenue}>50$ is stable across all contractions.

\section{Further experiments}
\label{sec:further-experiments}

\begin{figure*}
    \centering
    \includegraphics[width=\textwidth]{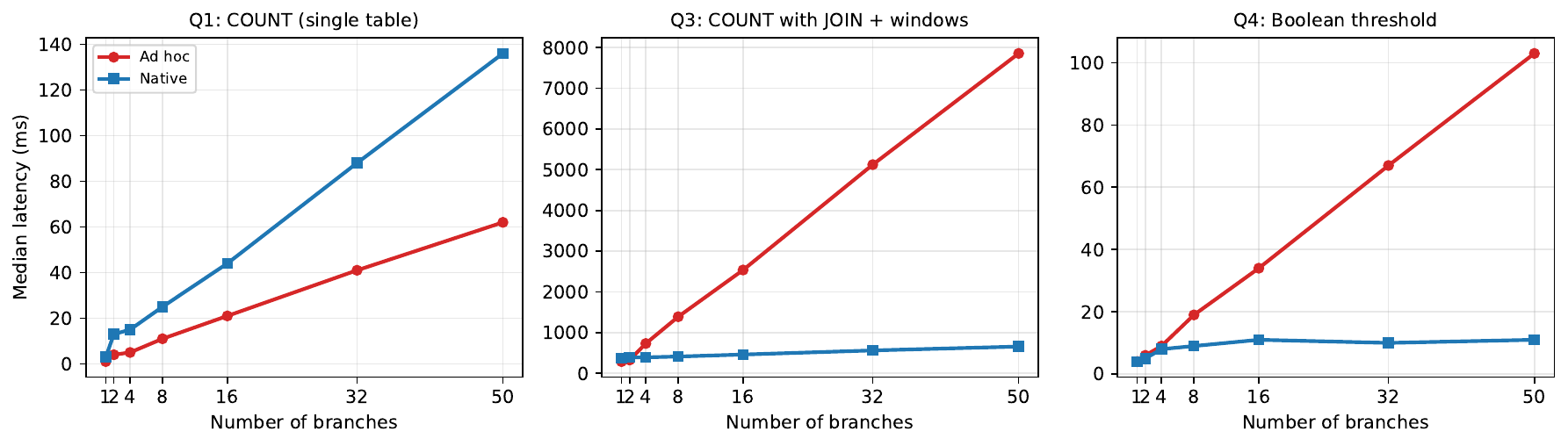}
    \caption{Median latency (ms) vs.\ number of branches. Left: Q1, both engines scale linearly but the native engine's unified plan carries coordination overhead (hash-repartitioning for \texttt{GROUP BY}). Center: Q3, the ad hoc engine scales at ${\sim}160$\,ms/branch (recomputing window functions over the shared table), while the native engine stays nearly flat thanks to CTE reuse. Right: Q4, the native engine's short-circuit optimization achieves near-constant ${\sim}11$\,ms latency.}
    \label{fig:bench-latency}
\end{figure*}

Table~\ref{tab:bench-results} and Fig.~\ref{fig:bench-latency} give the full data behind Section~\ref{sec:experiments}. The workload uses 54 branch variants of the predictions table (2.5M rows each) and a shared 3M-row dimension table as local Parquet files, with $B \in \{1,2,4,8,16,32,50\}$. Each configuration is run 5 times after 1 warmup; we report median wall-clock time (Apple M3, 36\,GB RAM).

\begin{table}[h]
  \small
  \caption{Median latency (ms) by engine and branch count.}
  \label{tab:bench-results}
  \begin{tabular}{@{}llrrrrrrr@{}}
    \toprule
    Q & Engine & 1 & 2 & 4 & 8 & 16 & 32 & 50 \\
    \midrule
    Q1 & Ad hoc   & \textbf{1} & \textbf{4} & \textbf{5} & \textbf{11} & \textbf{21} & \textbf{41} & \textbf{62} \\
    Q1 & Native   & 3 & 13 & 15 & 25 & 44 & 88 & 136 \\
    \midrule
    Q3 & Ad hoc   & \textbf{288} & 327 & 732 & 1390 & 2537 & 5125 & 7856 \\
    Q3 & Native   & 366 & \textbf{389} & \textbf{392} & \textbf{412} & \textbf{461} & \textbf{562} & \textbf{657} \\
    \midrule
    Q4 & Ad hoc   & 4 & 6 & 9 & 19 & 34 & 67 & 103 \\
    Q4 & Native   & \textbf{4} & \textbf{5} & \textbf{8} & \textbf{9} & \textbf{11} & \textbf{10} & \textbf{11} \\
    \bottomrule
  \end{tabular}
\end{table}

\end{document}